\documentclass[aps,prl,twocolumn,numerical,superscriptaddress,nofootinbib,showpacs,showkeys,longbibliography]{revtex4-1}

\usepackage{graphicx,color}
\usepackage{amsmath,amssymb,amsfonts}
\usepackage[colorlinks=true,linkcolor=blue,plainpages=false]{hyperref}

\newcommand{\be}{\begin{equation}}
\newcommand{\ee}{\end{equation}}
\newcommand{\ba}{\begin{eqnarray}}
\newcommand{\ea}{\end{eqnarray}}
\newcommand{\ban}{\begin{eqnarray*}}
\newcommand{\ean}{\end{eqnarray*}}

\def\v2{\mbox{$v_2$}}

\def\sqrtsNN{\mbox{$\sqrt{s_{\mathrm{NN}}}$}}

\begin{document}

\title{A New Correlator to Detect and Characterize the Chiral Magnetic Effect}
\medskip

\author{Niseem~Magdy} 
\email{niseemm@gmail.com}
\affiliation{Department of Chemistry, State University of New York, Stony Brook, New York 11794, USA}

\author{Shuzhe Shi}
\affiliation{Physics Department and Center for Exploration of Energy and Matter,
Indiana University, 2401 N Milo B. Sampson Lane, Bloomington, IN 47408, USA.}

\author {Jinfeng Liao} 
\affiliation{Physics Department and Center for Exploration of Energy and Matter,
Indiana University, 2401 N Milo B. Sampson Lane, Bloomington, IN 47408, USA.}

\author {N.~Ajitanand} 
\affiliation{Department of Chemistry, State University of New York, Stony Brook, New York 11794, USA}

\author{Roy~A.~Lacey} 
\email{Roy.Lacey@stonybrook.edu}
\affiliation{Department of Chemistry, State University of New York, Stony Brook, New York 11794, USA}

\date{\today}

\begin{abstract}
  A charge-sensitive in-event correlator is proposed and tested for its efficacy 
to detect and characterize charge separation associated with the 
Chiral Magnetic Effect (CME) in heavy ion collisions.
Tests, performed with the aid of two reaction models, indicate discernible responses 
for background- and CME-driven charge separation, relative
to the second- ($\Psi_{2}$) and third-order ($\Psi_{3}$) event planes, which could 
serve to identify the CME. The tests also indicate a degree of sensitivity which 
would enable robust characterization of the CME via Anomalous Viscous Fluid Dynamics (AVFD) 
model comparisons. 
	
%
\end{abstract}

\maketitle

High-energy nuclear collisions at the Relativistic Heavy Ion Collider (RHIC) and the Large Hadron Collider (LHC) 
can result in the creation of a plasma composed of strongly coupled chiral quarks and gluons or the 
Quark-Gluon Plasma (QGP). Topological transitions such as sphalerons~\cite{Klinkhamer:1984di,Arnold:1987mh}, 
which occur frequently in the QGP \cite{Moore:2010jd,Mace:2016svc}, can induce a net axial charge 
asymmetry of the chiral quarks which fluctuate from event to event.   
In the presence of the strong electromagnetic $\vec{B}$-fields created in the same  
collisions, this chiral anomaly is predicted to convert into an electric 
current which produces a final-state charge separation known as the Chiral Magnetic Effect 
(CME) \cite{Kharzeev:2004ey,Kharzeev:2007tn,Kharzeev:2007jp,Fukushima:2008xe,Kharzeev:2010gr,Jiang:2016wve}. 
For recent reviews, see e.g. \cite{Kharzeev:2013ffa,Liao:2014ava,Kharzeev:2015znc}.

The electric current  $\vec{J}_Q$, created along the $\vec{B}$-field, stems from anomalous chiral 
transport of the chiral fermions in the QGP:
\begin{eqnarray} \label{eq_cme}
\vec{J}_Q &=& \sigma_5 \vec{B},\ \ \ \ \sigma_5 = \mu_5 \frac{Q^2}{4\pi^2},
\end{eqnarray} 
where $\sigma_5$ is the chiral magnetic conductivity, 
$\mu_5$ is the chiral chemical potential that quantifies the axial charge 
asymmetry or imbalance between right-handed and 
left-handed quarks in the plasma, and $Q$ is the  quark electric 
charge \cite{Fukushima:2008xe,Son:2009tf,Zakharov:2012vv,Fukushima:2012vr}. 
Thus, experimental observation of its associated charge separation, could 
provide crucial insights on anomalous transport and the interplay 
of chiral symmetry restoration, axial anomaly, and gluonic topology in the QGP.

The $\vec{B}$-field, which is strongly 
time-dependent~\cite{Skokov:2009qp,McLerran:2013hla,Tuchin:2014iua}, is generated perpendicular 
to the reaction plane ($\mathrm{\Psi_{RP}}$) defined by the impact parameter and the beam axis. 
Consequently, CME-driven charge separation can be identified and characterized via the first 
$P$-odd sine term ($\mathrm{a_{1}}$) in a Fourier decomposition of the 
charged-particle azimuthal distribution~\cite{Voloshin:2004vk}:
\begin{eqnarray}\label{eq:1}
\mathrm{\frac{dN^{ch}}{d\phi} \propto [1 + 2\sum_{n} v_{n} \cos(n \Delta\phi) + a_n sin(n \Delta\phi)  + ...]}
\end{eqnarray}
where $\mathrm{\Delta\phi = \phi -\Psi_{RP}}$ gives the particle azimuthal angle
with respect to the reaction plane angle, and $\mathrm{v_{n}}$ and $\mathrm{a_{n}}$ denote the
coefficients of $P$-even and $P$-odd Fourier terms, respectively. 
The second-order event plane, $\Psi_{2}$, determined by the maximal particle
density in the elliptic azimuthal anisotropy and the beam axis, is usually employed as 
a proxy for $\mathrm{\Psi_{RP}}$ in experimental measurements. 
Here, it is noteworthy that the third-order event plane, $\Psi_{3}$, can not be used 
to detect CME-driven charge separation, since there is little, if any, correlation 
between $\mathrm{\Psi_{RP}}$ and $\Psi_{3}$.   
The event-by-event fluctuations contribute to an event-wise de-correlation between 
the magnetic field direction imposed by $\Psi_{\bf \rm RP}$, and the 
orientation of $\Psi_{2}$ imposed by the bulk collision geometry~\cite{Bloczynski:2012en}.
The dispersion of $\mathrm{\Psi_{2}}$ about $\Psi_{\bf \rm RP}$ reduces the 
magnitude of $\mathrm{a_{1}}$, which depends on both the initial axial charge 
and the time evolution of the magnetic field (c.f. Eq.~\ref{eq_cme}). The latter are 
both not well constrained theoretically.

The charge-dependent correlator, $\gamma_{\alpha\beta}$,  has been widely used 
at  RHIC \cite{Abelev:2009ac,Abelev:2009ad,Adamczyk:2013hsi,Adamczyk:2013kcb,Adamczyk:2014mzf,Tribedy:2017hwn,Zhao:2017nfq}  
and the LHC \cite{Abelev:2012pa,Khachatryan:2016got} in ongoing attempts to 
identify and quantify CME-driven charge separation:
\be
\gamma_{\alpha\beta} = \left\langle \cos\big(\phi_\alpha^{(\pm)} +
\phi_\beta^{(\pm)} -2 \Psi_{\rm 2}\big) \right\rangle,
\label{eq:2}
\ee
where $\phi_\alpha,\phi_\beta$ denote the azimuthal emission angles for like-sign ($++ \text{or} --$) 
and unlike-sign ($+\,-$) particle pairs. A charge-dependent azimuthal correlation, qualitatively 
consistent with the expectation for CME-driven charge separation, has been observed in these 
measurements. However, they remain inconclusive because of several identified 
sources of background correlations that can account for most, if not all, of the 
measurements~\cite{Wang:2009kd,Bzdak:2010fd,Schlichting:2010qia,Muller:2010jd,Liao:2010nv}.
A recent cause for pause, is the observation that the charge-dependent azimuthal correlations
for p+Pb and Pb+Pb collisions, have nearly identical values for similar multiplicity 
selections~\cite{Khachatryan:2016got}. This poses a significant challenge for the use of 
the $\gamma_{\alpha\beta}$ correlator in such measurements, because CME-induced charge separation is 
predicted to be negligible in p+Pb collisions. That is, the absence of a strong correlation  
between the orientation of the $\Psi_2$ plane and the $\vec{B}$-field 
in p+Pb collisions, should result in very little, if any,
CME-driven charge separation~\cite{Khachatryan:2016got,Belmont:2016oqp,Laceytalk2017}.

To a large extent, the present ambiguity between background- and CME-driven charge separation, 
stems from the fact that the $\gamma_{\alpha\beta}$ correlator gives the same qualitative 
response to both. Thus, new measurements and improved data analysis methodologies, designed to  
suppress or separate background contributions from genuine CME-driven charge separation, are 
required for robust identification and characterization of the CME \cite{Sirunyan:2017quh}.

In this work we present and validate the response of a new charge-sensitive 
correlator, specifically designed to give distinct discernible responses for 
background- and CME-driven charge separation relative to the $\Psi_{2}$ and $\Psi_{3}$ event planes. 
The tests are performed with A Multi-Phase Transport Model (AMPT) \cite{Lin:2004en} and the 
Anomalous Viscous Fluid Dynamics (AVFD) model \cite{Jiang:2016wve}. 
Both models are known to give a good representation of the experimentally measured 
particle yields, spectra, flow, etc. Therefore, they both can provide a good estimate 
of the magnitude and nature of the purely background-driven charge separation signal 
expected in the data samples collected at RHIC and the LHC.

Anomalous transport from the CME, is also implemented in the AVFD model \cite{Jiang:2016wve}. 
This important feature, facilitates our study of the correlators' response 
to the combined influence of the backgrounds and an input CME-driven charge separation signal.
The model uses Monte Carlo Glauber initial conditions to simulate the evolution 
of fermion currents in the QGP, on top of the bulk fluid evolution implemented in 
the VISHNU hydrodynamic code, followed by a URQMD hadron cascade stage. 
A time-dependent magnetic field $B(\tau) = \frac{B_0}{1+\left(\tau / \tau_B\right)^2}$,
acting in concert with a nonzero initial axial charge density, is used to generate 
a CME current (embedded in the fluid dynamical equations) leading to a charge separation 
along the magnetic field. The peak values $B_0$, obtained from event-by-event 
simulations~\cite{Bloczynski:2012en}, are used with a relatively conservative 
lifetime $\tau_B=0.6$ fm/c. For the initial axial charge density arising from gluonic 
topological charge fluctuations, we adopt the commonly used estimate based on the strong 
chromo-electromagnetic fields in the early-stage glasma. A More in-depth account of the 
implementation of the AVFD model can be found in Refs.~\cite{Jiang:2016wve} and \cite{Shi:2017cpu}.

The new correlators $R_{\Psi_m}(\Delta S)$, are constructed for each  
event plane $\Psi_m$, as the ratio:
\be
R_{\Psi_m}(\Delta S) = C_{\Psi_m}(\Delta S)/C_{\Psi_m}^{\perp}(\Delta S), \, m=2,3 ,
\label{eq:4}
\ee
where $C_{\Psi_m}(\Delta S)$ and $C_{\Psi_m}^{\perp}(\Delta S)$ are correlation functions
designed to quantify charge separation $\Delta S$, parallel and perpendicular (respectively) to 
the $\vec{B}$-field, i.e., perpendicular and parallel (respectively) to $\mathrm{\Psi_{RP}}$. 
Since CME-driven charge separation occurs only along the $\vec{B}$-field and 
$\Psi_2$ and $\mathrm{\Psi_{RP}}$ are strongly correlated, $C_{\Psi_2}(\Delta S)$ measures 
both CME- and backgrond-driven charge separation. In contrast, $C_{\Psi_2}^{\perp}(\Delta S)$ measures 
only background-driven charge separation. 
The absence of a strong correlation between the orientation of the $\Psi_3$ plane 
and the $\vec{B}$-field, also renders $C_{\Psi_3}(\Delta S)$ and $C_{\Psi_3}^{\perp}(\Delta S)$
insensitive to a CME-driven charge separation. However, they provide crucial 
insight on the relative importance of background-driven charge separation as discussed below.

The correlation functions used to quantify charge separation parallel to the $\vec{B}$-field, 
are constructed from the ratio of two distributions \cite{Ajitanand:2010rc}: 
\be
C_{\Psi_{m}}(\Delta S) = \frac{N_{\text{real}}(\Delta S)}{N_{\text{Shuffled}}(\Delta S)}, \, m=2,3,
\label{eq:5}
\ee
where $N_{\text{real}}(\Delta S)$ is the distribution over events, of charge separation 
relative to the $\Psi_m$ planes in each event:
\be
\Delta S = \frac{{\sum\limits_1^p {\sin (\frac{m}{2}\Delta {\varphi_{m} })} }}{p} - 
\frac{{\sum\limits_1^n {\sin (\frac{m}{2}\Delta {\varphi_{m}  })} }}{n},
\label{eq:7}
\ee
where $n$ and $p$ are the numbers of negatively- and positively charged hadrons in an event, 
$\Delta {\varphi_{m}}= \phi - \Psi_{m}$ and $\phi$ is the 
azimuthal emission angle of the charged hadrons. The $N_{\text{Shuffled}}(\Delta S)$ distribution
is similarly obtained from the same events, following random reassignment (shuffling) of the charge of 
each particle in an event. This procedure ensures identical properties for the 
numerator and the denominator in Eq.~\ref{eq:5}, except for the charge-dependent correlations 
which are of interest.

The correlation functions $C_{\Psi_{m}}^{\perp}(\Delta S)$, used to quantify charge separation 
perpendicular to the $\vec{B}$-field, are constructed with the same procedure outlined 
for $C_{\Psi_{m}}(\Delta S)$, but with $\Psi_{m}$ replaced by $\Psi_{m}+\pi/m$.
This $\pi/m$ rotation of the event plane, guarantees that a possible CME-driven charge 
separation does not contribute to these correlation functions. 

The correlator $R_{\Psi_2}(\Delta S) = C_{\Psi_2}(\Delta S)/C_{\Psi_2}^{\perp}(\Delta S)$, 
gives a measure of the magnitude of charge separation parallel to the $\vec{B}$-field (perpendicular to $\Psi_2$),
relative to that for charge separation perpendicular to the $\vec{B}$-field (parallel to $\Psi_2$).
Since the CME occurs along the $\vec{B}$-field, correlations dominated by CME-driven charge separation 
should result in concave-shaped distributions having widths that reflect the magnitude $\mathrm{a_1}$ of the 
charge separation (cf. Eq.~\ref{eq:1}). That is, the stronger the CME-driven charge separation, the narrower 
the $R_{\Psi_2}(\Delta S)$ distribution. In contrast, the correlator 
$R_{\Psi_3}(\Delta S) = C_{\Psi_3}(\Delta S)/C_{\Psi_3}^{\perp}(\Delta S)$ would be insensitive 
to this CME-driven charge separation, due to the absence of a strong correlation between the orientation 
of the $\Psi_3$ plane and the $\vec{B}$-field.

For background-driven charge separation, similar patterns are to be expected for 
both the $R_{\Psi_2}(\Delta S)$ and $R_{\Psi_3}(\Delta S)$ distributions. Note as well,
that such patterns could be convex- or concave-shaped \cite{Bozek:2017plp}, depending on the detailed 
nature of the background-driven correlations. Therefore, in addition to an observed concave-shaped 
distribution for $R_{\Psi_2}(\Delta S)$, an observed difference between the distributions for 
$R_{\Psi_2}(\Delta S)$ and $R_{\Psi_3}(\Delta S)$ is essential for CME identification and 
characterization.

The magnitude of a CME-driven charge separation is reflected in the width of the concave-shaped 
distribution for $R_{\Psi_2}(\Delta S)$, which is also influenced by particle number 
fluctuations and the resolution of $\Psi_2$. 
That is, stronger CME-driven signals lead to narrower concave-shaped distributions (smaller widths), 
which are made broader by particle number fluctuations and poorer event-plane resolutions. 
The influence of the particle number fluctuations can be minimized by scaling $\Delta S$ 
by the width $\mathrm{\sigma_{\Delta_{Sh}}}$ of the distribution 
for $N_{\text{shuffled}}(\Delta S)$ {\em i.e.}, $\Delta S^{'} = \Delta S/\mathrm{\sigma_{\Delta_{Sh}}}$. 
Similarly, the effects of the event plane resolution can be accounted for by scaling 
$\Delta S^{'}$ by the resolution factor $\mathrm{\delta_{Res}}$, {\em i.e.}, 
$\Delta S^{''}= \Delta S^{'}/\mathrm{\delta_{Res}}$, where $\mathrm{\delta_{Res}}$ is the event 
plane resolution. The efficacy of these scaling factors have been confirmed via detailed 
simulation studies, as well as with actual data.  

\begin{figure}[t]
\includegraphics[width=1.0\linewidth, angle=-90]{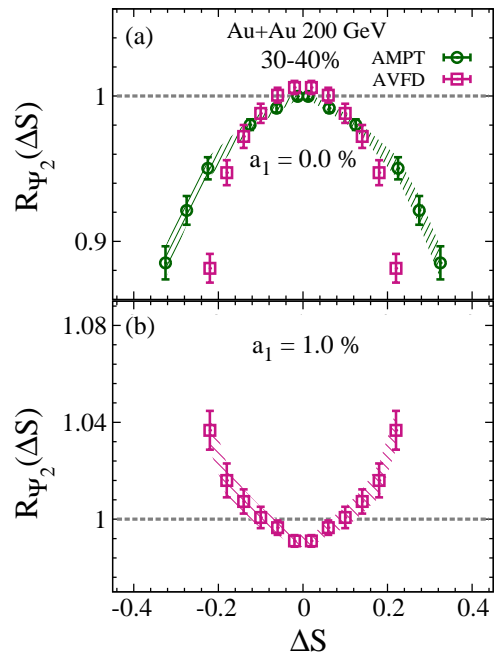}
\caption{ Comparison of the $R(\Delta S)$ correlators for (a) background-driven charge 
separation ($a_{1} = 0$) in 30-40\% Au+Au collisions ($\sqrtsNN~=~200$ GeV) 
obtained with the AMPT and AVFD models, and (b) the combined effects of background- and 
CME-driven ($a_{1} = 1.0\%$) charge separation in Au+Au collisions obtained with 
the AVFD model at the same centrality and beam energy. 
}
\label{fig1} 
\end{figure}

\begin{figure}[t]
\includegraphics[width=1.0\linewidth]{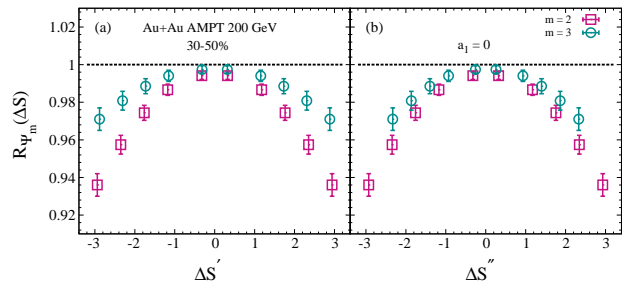} 
\caption{ Comparison of the $R_{\Psi_m}(\Delta S^{'})$ (a) and $R_{\Psi_m}(\Delta S^{''})$ (b) 
correlators for background-driven charge separation ($a_{1} = 0$) 
in 30-50\% Au+Au collisions ($\sqrtsNN~=~200$ GeV) obtained with the AMPT model.
} 
\label{fig2} 
\end{figure}

\begin{figure}[t]
\includegraphics[width=0.70\linewidth, angle=-90]{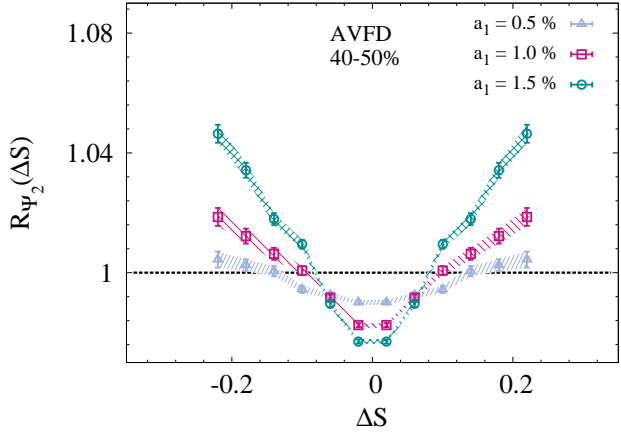}
\vskip -0.10in
\caption{ Comparison of the $R_{\Psi_2}(\Delta S)$ correlator for different input charge separation 
signals characterized by $\mathrm{a_{1}}$, in 40-50\% central Au+Au ($\sqrtsNN~=~200$ GeV)
events obtained with AVFD model.
} 
\label{fig3} 
\end{figure}

\begin{figure}[t]
%
\includegraphics[width=0.75\linewidth, angle=-90]{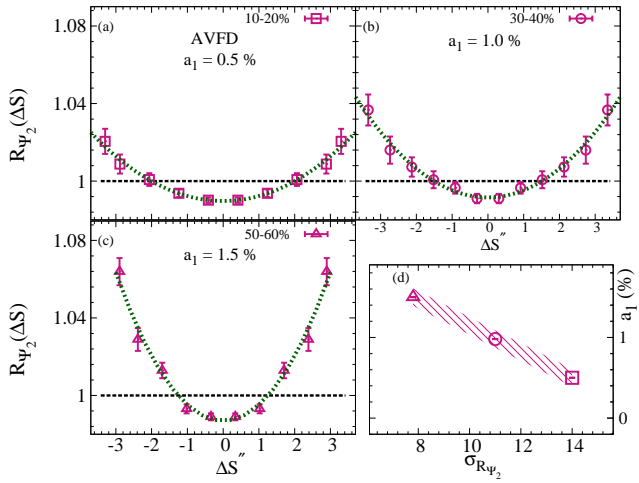}
\caption{ Comparison of the $R_{\Psi_2}(\Delta S^{''})$ correlator obtained from the 
analysis of (a) 10-20\%, (b) 30-40\% and (c) 50-60\% central AVFD events 
for Au+Au collisions at $\sqrtsNN~=~200$ GeV. The curves represent Gaussian fits 
to the distributions; (d) $\mathrm{a_{1}}$ vs. $\mathrm{\sigma_{R_{\Psi_2}}}$ for 
the fits indicated indicated in (a), (b) and (c).
} 
\label{fig4} 
\vspace{-0.1in}
\end{figure}

\begin{figure}[t]
\includegraphics[width=0.65\linewidth, angle=-90]{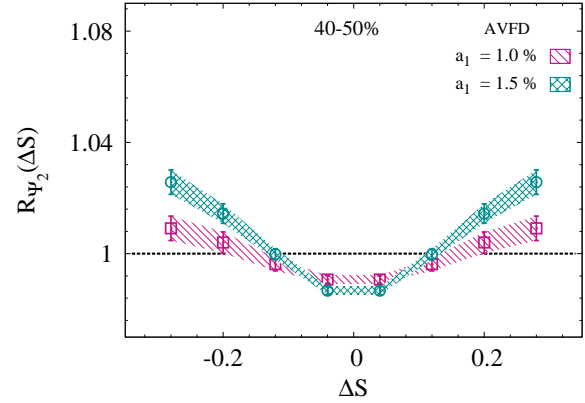}
\vskip -0.10in
\caption{ Comparison of the $R_{\Psi_2}(\Delta S)$ correlators obtained from
AVFD events with $\mathrm{a_{1}} = 1\%$ and 1.5\%. The AVFD results 
are generated with an event plane resolution and $p_T$ and $\eta$ cuts, 
similar to the experimental ones~\cite{Laceytalk2017}. 
} 
\label{fig5} 
\vspace{-0.1in}
\end{figure}

Simulated events from both the AMPT and AVFD models were used to study the response, as well as 
the efficacy of the $R_{\Psi_m}(\Delta S)$ correlators. Representative results from these 
studies are summarized in Figs.~\ref{fig1}~-~\ref{fig5}. 

The response of the correlator to 
background- and CME-driven charge separation is illustrated in Fig.~\ref{fig1}. 
Panel (a) indicates that the $R_{\Psi_{2}}(\Delta S)$ correlator 
exhibits a convex-shaped distribution for the background-driven 
($a_{1} = 0$) charge separation in both models, albeit with some model 
dependence for the magnitudes. Note that these background-driven distributions 
are not required to be convex-shaped \cite{Bozek:2017plp} and are specific to these models. 
Panel (b) shows that the introduction of a modest input CME-driven charge 
separation ($a_{1} = 1.0\%$) in the same AVFD events, results in a change from convex-shaped
to a concave-shaped distribution for $R_{\Psi_{2}}(\Delta S)$. This change reflects 
the influence of the CME-driven charge separation in the AVFD model.
These patterns contrast with those of the the $\gamma_{\alpha\beta}$ correlator, which 
was observed to give the same qualitative response to both background-driven and CME-driven charge 
separation in AMPT model simulations \cite{Ma:2011uma}. 

%
%
%

%
%
%
%
%

%
%
%
%

%
%
%

%
%
%


Figure \ref{fig2} show background-driven charge separation distributions for 
both $R_{\Psi_{2}}(\Delta S)$ and $R_{\Psi_{3}}(\Delta S)$, obtained with the 
AMPT model. Panels (a) and (b) show distributions which are corrected for number 
fluctuations ($\Delta S^{'}$) and the combined effects of number fluctuations and 
event plane resolution ($\Delta S^{''}$) respectively.
Fig.~\ref{fig2}(b) indicate the expected similarity between the shape and widths 
for $R_{\Psi_{2}}(\Delta S^{''})$ and $R_{\Psi_{3}}(\Delta S^{''})$. 
This similarity is especially important since $R_{\Psi_{3}}(\Delta S)$ 
is insensitive to CME-driven charge separation.
Thus, a discernible difference in the response for $R_{\Psi_{2}}(\Delta S^{''})$ 
and $R_{\Psi_{3}}(\Delta S^{''})$ constitutes a crucial and necessary requirement 
for unambiguous identification and characterization of CME-driven charge separation. 
In the same vein, $R_{\Psi_{2}}(\Delta S)$ would not be expected to show 
a significant concave-shaped response in p(d)+A collisions, due to the absence 
of a strong correlation between the orientation of the $\Psi_2$ plane and the 
$\vec{B}$-field in these collisions~\cite{Khachatryan:2016got,Belmont:2016oqp,Laceytalk2017}.

The sensitivity of the $R_{\Psi_{2}}(\Delta S)$ correlator to varying degrees of input CME-driven charge 
separation (characterized by $\mathrm{a_{1}}$) at a fixed collision centrality, is 
shown in Fig.~\ref{fig3}. Note that for a fixed centrality, a change in the value of 
$\mathrm{a_{1}}$ is tantamount to a change in the input value of the initial chiral 
anomaly in AVFD. Note as well that, for a fixed centrality, the event plane resolution is 
the same for events generated with different values of $\mathrm{a_{1}}$. 
A concave-shaped distribution can be observed 
in each case, confirming the presence of the input CME-driven signals. The amplitudes of these 
distributions also track with the magnitude of $\mathrm{a_{1}}$, indicating that the 
$R_{\Psi_{2}}(\Delta S)$ correlator is not only suited for CME-driven signal identification, but also 
for signal characterization. 

The sensitivity of the $R_{\Psi_{2}}(\Delta S)$ correlator to the influence of the 
$\vec{B}$-field in AVFD, can also be studied via the centrality dependence 
of $R_{\Psi_{2}}(\Delta S)$. Figs.~\ref{fig4}(a), (b) and (c) show the correlator 
distributions for $10-20\%$, $30-40\%$ and $50-60\%$ central Au+Au collisions respectively. 
For these plots, we have scaled $\Delta S$ to account for the difference in the 
associated number fluctuations and event plane resolution ($\Delta S^{''}$). 
The concave-shaped distribution, apparent in each panel, confirms the input 
CME-driven signal in each case. The widths of these distributions $\mathrm{\sigma_{R_{\Psi_2}}}$, 
also reflect the increase of $\mathrm{a_{1}}$ as collisions become more peripheral (panel (d)). 
This confirms the trend expected for the magnitude of the $\vec{B}$-field 
with collision centrality. The implied sensitivity of $R_{\Psi_{2}}(\Delta S)$ to 
the $\vec{B}$-field, could also provide an independent approach to the detection 
and quantification of the CME in isobaric collision.

For model comparisons to actual experimental data, it is also necessary to impose the 
experimental cuts ($\left| \eta \right| \sim 0.8$ and $p_T \geq 0.35$~GeV/$c$~\cite{Laceytalk2017}), 
as well as account for any difference between 
the experimental and simulated event plane resolution. Fig.~\ref{fig5} compares the results 
from AVFD for two values of $\mathrm{a_{1}}$, when the experimental cuts and the 
event plane resolution are taken into account.  These $R(\Delta S)$ distributions are 
still concave-shaped, albeit with smaller amplitudes than the corresponding distributions
shown in Fig.~\ref{fig3}, due to the effects of the kinematic cuts and the event plane resolution. 
A rudimentary comparison of these distributions to preliminary 
STAR Au+Au data for the same centrality and beam energy \cite{Laceytalk2017} 
shows good agreement between the data and AVFD results for $a_{1} = 1\%$, suggesting  
the presence of a small CME-driven charge separation in 40-50\% central Au+Au collisions 
at $\sqrtsNN~=~200$ GeV. 

In summary, we have presented new $R_{\Psi_{m}}(\Delta S)$ correlators which are well suited for 
studying the CME. Validation tests, performed with the AMPT and AVFD models, indicate that the 
correlator gives discernible responses for background- and CME-driven charge 
separation, which could allow unambiguous identification of the CME via $R_{\Psi_{2}}(\Delta S)$
and $R_{\Psi_{3}}(\Delta S)$ measurements. 
The tests also indicate a degree of sensitivity which would enable a robust characterization 
of experimental CME-driven charge separation signals with magnitudes comparable to those 
currently simulated in the AVFD model. An initial comparison 
of the correlators obtained from preliminary data and AVFD calculations, suggests
the presence of a CME-driven charge separation in 40-50\% central Au+Au 
collisions at $\sqrtsNN~=~200$ GeV. A vigorous effort is currently underway 
to extract the experimental and theoretical differential $R_{\Psi_{m}}(\Delta S)$ correlators 
for different systems and energies, to characterize the CME in RHIC and LHC 
collisions. The $R_{\Psi_{m}}(\Delta S)$ correlators also provide an independent 
approach to the detection and quantification of the CME in the upcoming 
isobaric collision experiments at RHIC.

\section*{Acknowledgments}
\begin{acknowledgments}
This research is supported by the US Department of Energy, Office of Science, Office of Nuclear Physics, 
under contract DE-FG02-87ER40331.A008  (NM, NA and RL) and by the National Science Foundation 
under Grant No. PHY-1352368 (SS and JL).  
The AVFD study is based upon work supported by the U.S. Department of Energy, 
Office of Science, Office of Nuclear Physics, within the framework of 
the Beam Energy Scan Theory (BEST) Topical Collaboration.
\end{acknowledgments}
%
%
\bibliography{lpvpub} 
\end{document}